\documentstyle[12pt,epsf]{article}
\renewcommand{\bar}[1]{\overline{#1}}

\def\Vec#1{\hbox{\boldmath$#1$\unboldmath}}

\begin{document}


\begin{center}
{\Large \bf
Connection between Distribution \\
and Fragmentation Functions}

\vspace{2cm}

{\bf Vincenzo Barone$^{a}$, Alessandro Drago$^{b}$,
and Bo-Qiang Ma$^{c}$} \\

\vspace{0.5cm}

\baselineskip=14pt

 $^{a}$Dipartimento di Fisica Teorica,
Universit\`a  di Torino, \\
INFN, Sezione di Torino, 10125 Torino,
Italy, \\
and D.S.T.A., Universit{\`a} ``A.~Avogadro'', 15100 Alessandria,
Italy

\vspace{0.2cm}

 $^{b}$Dipartimento di Fisica,
Universit\`a  di Ferrara, \\
and INFN, Sezione di Ferrara, 44100 Ferrara,
Italy

\vspace{0.2cm}

$^{c}$CCAST (World Laboratory), P.O.~Box 8730, Beijing
100080, China, \\
Department of Physics, Peking University,
Beijing 100871, China\footnote{Mailing address.},\\
and Institute of Theoretical Physics, Academia Sinica,
Beijing 100080, China

\end{center}


\baselineskip=14pt

\vspace{2cm}
\begin{center} {\large \bf Abstract}

\end{center}

We show that the quark fragmentation function $D(z)$ and the quark
distribution function $q(x)$ are connected in the $z \to 1$ limit
by the approximate relation $D(z)/z \simeq  q(2-1/z)$, where both
quantities are in their physical regions. Predictions for proton
production in inelastic $e^+ e^-$ annihilation, based on the new
relation and standard parametrizations of quark distribution
functions, are found to be compatible with the data.

\vfill

\centerline{PACS numbers: 13.65.+i, 11.80.-m, 12.39.-x, 13.60.Hb}

\vfill
\centerline{Published in Phys.~Rev.~C 62 (2000) 062201(R) 
as a rapid communication}
\vfill
\newpage

\baselineskip=18pt


Inclusive deep inelastic scattering (DIS)  and
hadron production in $e^+e^-$ inelastic annihilation (IA)
are important
sources of information on the
structure of hadrons.
By simple crossing it is possible
to relate the structure functions of these
two processes \cite{DLY}.
However,  the relation so
obtained (known as the Drell-Levy-Yan relation) is of little use,
as it connects the structure functions in the {\em physical} region
of one process to the structure functions in the {\em unphysical}
region of the other process.

It would obviously be important
to have a relation between the DIS and IA
structure functions, taken both in their physical regions.
We could then exploit the accurate information we already possess
on the quark distribution functions of the nucleon
to predict the quark fragmentation functions, which
are still poorly known, or {\it vice versa}, by measuring
the fragmentation functions of hadrons which cannot
be used as DIS targets
we could predict the quark densities inside those hadrons.
An example which immediately
comes to mind is the $\Lambda$:
the quark dynamics inside this hyperon
is highly relevant for our understanding
of the spin and  flavor structure
of hadrons \cite{Bur93,Ma99,Ma00}.

As a matter of fact, a
relation connecting the structure functions of DIS
and IA in their physical regions does  exist
in the literature.
It is the so-called
Gribov-Lipatov ``reciprocity" relation.
As we shall see, this relation, in its commonly used form,
has no real
justification.  Moreover it is not supported
by phenomenological evidence.
The purpose of this paper is to derive
another, well founded, relation
connecting DIS and IA, and to show that this relation,
within its range of validity, is in
good agreement with the existing data.

The DIS cross section is written in terms of two
structure functions $F_1(x, Q^2)$ and $F_2(x, Q^2)$,
where $x = Q^2/ 2 p \cdot q$ is the Bjorken variable
and $Q^2 = - q^2$ is the momentum transfer squared.
Two analogous quantities appear
in the IA cross
section: they
are denoted by $\bar F_1(z, Q^2)$ and
$\bar F_2(z, Q^2)$, where now $z = 2 p \cdot q/ Q^2$ and
$ Q^2 = q^2$ is the center-of-mass energy squared.
In leading order QCD
the Callan-Gross relations connect $F_1 \, (\bar F_1)$
to $F_2 \, (\bar F_2)$ as
\begin{equation}
F_2(x, Q^2) = 2x\,  F_1(x, Q^2) \,,
\label{CG2}
\end{equation}
\begin{equation}
2 z \, \bar F_1(z, Q^2) = - z^2 \, \bar F_2(z, Q^2) \,,
\label{CG}
\end{equation}
and, at the same order, the structure functions
 can be expressed
in terms of the distribution functions $q_{a, \bar a}$
and  the fragmentation functions
 $D_{a, \bar a}$ as
\begin{equation}
2 F_1(x, Q^2) = \frac{1}{x} F_2(x, Q^2) = \sum_a e_a^2
\,  [q_a(x, Q^2) + q_{\bar a}(x, Q^2)]\;,
\label{distr}
\end{equation}
\begin{equation}
2 z \bar F_1 (z, Q^2) = - z^2 \bar F_2 (z, Q^2)
= 3 \, \sum_a e_a^2 [ D_a(z, Q^2) + D_{\bar a}(z, Q^2) ]\,,
\label{fragm}
\end{equation}
where the sums run over all flavors (the factor 3 comes
from a sum over colors).

The traditional form of the Gribov-Lipatov relation
reads \cite{Fis72}
\begin{equation}
\begin{array}{c}
z \bar{F}_1(z)=F_1(z), \\
 z^3 \bar{F}_2(z)=-F_2(z),
\end{array}
\label{GLRF}
\end{equation}
where $F_{1,2}(z)$ means that the DIS structure
functions are evaluated at $x = z$.
Phenomenological
tests of this relation have been carried out
\cite{DASP,Kon93,Pet99} and it turns out that
the IA structure functions predicted by Eq.~(\ref{GLRF})
undershoot the data.
But the main shortcoming of (\ref{GLRF})
is its uncertain theoretical status.
In fact, what Gribov and Lipatov proved in their classical
papers \cite{GLR} is that
  the non-singlet splitting functions
for DIS and IA are equal at leading order
(for a detailed and clear discussion see \cite{Altarelli})
\begin{equation}
\bar  P_{qq}^{(0)}(z) = P_{qq}^{(0)} (z)\;.
\label{GL}
\end{equation}
Thus Eq.~(\ref{GLRF}) is true at leading order  if
one assumes that at a nonperturbative scale $\mu^2$ the
input distribution and fragmentation functions are
related to each other by
\begin{equation}
3 D(z) = q (z)\;.
\label{GLRF2}
\end{equation}
This relation is  unjustified.
We will now show that the true  nonperturbative
relation existing between $q(x)$ and $D(z)$
in the large-$z$ limit is the approximate
relation
\begin{equation}
\frac{1}{z} \, D(z) \simeq  \, q(2-1/z)\;.
\label{NR0}
\end{equation}
Since $2-1/z \simeq \frac{1}{1-(1-1/z)} =z$, as $z \to 1$,
Eq.~(\ref{NR0}) can be
 approximated further by
\begin{equation}
\frac{1}{z} \, D(z) \simeq  \, q(z)\;.
\label{GLRR}
\end{equation}
This relation was used as a phenomenological Ansatz in
 \cite{Ma00,Bro97}.

Let us start
from the general definition of the quark
distribution function \cite{Col82}
\begin{equation}
q(x)=\int\frac{\mathrm d \xi^-}{4\pi}e^{ixp^+\xi^-}
 \langle h(p)   \vert \overline{\psi}(0)\gamma^+ \psi(\xi^-)
\vert  h(p)  \rangle,
\end{equation}
for a hadron $h$ with mass $M$ and
 momentum $p=(p^0,\Vec{p})$. The light-cone
components are defined as $p^{\pm} = \frac{1}{2} (p^0 \pm p^3)$.
The normalization of the states is $\langle p \vert p' \rangle
= (2 \pi)^3 \, 2 p^+ \, \delta (p^+ - p'^+) \,
\delta^2 ({\Vec p}_{\perp} - {\Vec p'}_{\perp})$.
By inserting a complete set of intermediate states $\left|n\right>$
with momentum $p_n=(p_n^0, {\Vec p}_n)$ and mass $M_n$
and making use of the
translational invariance,
one obtains
\begin{equation}
q(x)=\frac{\sqrt{2}}{2}\sum_n
\int \frac{{\rm d}^4 p_n}{(2\pi)^3}\,
\delta(p_n^2-M^2_n) \,
\delta(p^+-x p^+-p^+_n)
\vert \langle n (p_n) \vert
\psi_+(0) \vert h(p) \rangle \vert^2,
\label{qxd}
\end{equation}
where $\psi_+=\frac{1}{2}\gamma^-\gamma^+\psi$ is the `good' component
the quark field operator.

Similarly, the fragmentation function of a quark
into an unpolarized hadron $h$ is defined as
\cite{Col82} (a sum over the final spin of the hadron
is performed)
\begin{eqnarray}
\frac{1}{z} \, D(z) &=&
\sum_{n}
\int\frac{{\rm d} \xi^-}{4\pi}e^{-ip^+\xi^-/z}
\, \int \frac{{\rm d}^4 p_n}{( 2 \pi)^3} \, \delta(p_n^2 -
M_n^2) \nonumber \\
&\times&
{\rm Tr} \{ \gamma^+  \langle 0 \vert  \psi (0)
\vert h(p), n(p_n)  \rangle  \langle h(p), n(p_n)
\vert \overline{\psi}(\xi^-)
 \vert 0  \rangle,
\end{eqnarray}
where the initial quark carries a light-cone momentum
$k^+=p^+/z$. The normalization
of $D(z)$ is such that $\sum_h \int {\rm d} z z D(z) =1$.
Using translational invariance  one obtains \cite{Thomas}
\begin{equation}
\frac{1}{z} \, D(z)=\frac{1}{2\sqrt{2}} \sum_n
\int \frac{\rm{d}^4 p_n}{(2\pi)^3} \, \delta(p_n^2-M^2_n)
\, \delta(p^+/z - p^+ -p^+_n)
\vert \langle 0 \vert
\psi_+(0) \vert h(p), n(p_n)  \rangle \vert^2.
\label{dzd}
\end{equation}
Crossing symmetry means
\begin{equation}
 \langle 0 \vert \psi_+(0) \vert h(p), n(p_n) \rangle =
 \langle \bar n (-p_n) \vert  \psi_+(0) \vert
h(p)  \rangle\,.
\label{cross}
\end{equation}
If we make the change $p_n \to - p_n$ in the integral
(\ref{dzd}) and use (\ref{cross}) we get
(remember that $ x = 1/z$)
\begin{equation}
\frac{1}{z} \, D(z) = q(1/z) \,.
\label{ndly}
\end{equation}
This relation, which connects the fragmentation function
in the physical region
$0 \le z \le 1$ to the quark distribution in the unphysical region
$x = 1/z \ge 1$, and {\it vice versa},
is equivalent to the Drell-Levy-Yan
relation \cite{DLY}.

A further step is needed in order to get the relation (\ref{NR0}).
What we have to do  is to establish a connection
between the physical
and the unphysical region of the
quark distribution function (or, equivalently, of the
fragmentation function).

In the physical region $0 \le x \le 1$ the $\delta$-function
in (\ref{qxd}) constrains $p_n^+ = (1-x) p^+$ to be
positive and hence selects positive energy
states in the sum over $n$. Eq.~(\ref{qxd}) can be
rewritten as
\begin{equation}
q(x)= \frac{\sqrt{2}}{4 \pi} \, \sum_n
\int \frac{{\rm d} {\Vec p}_n}{2 \vert p_n^0 \vert \, (2 \pi)^3} \,
\delta(p^+-x p^+-p^+_n) \,
\vert   \langle n (p_n) \vert
\psi_+(0) \vert h(p) \rangle \vert^2 \;,
\label{qxp}
\end{equation}
where $ p_n^0 = ( {\Vec p}_n^2 + M_n )^{1/2}$.
The $\delta$-function
allows the integration to be
simplified giving
\cite{noi}
\begin{equation}
\int {\rm d} {\Vec p}_n \, \delta [p^+ - xp^+ -p_n^+]
=2 \pi \int_{p_{min}}^{\infty} {\rm d} \vert {\Vec p}_n \vert \,
\vert {\Vec p}_n \vert,
\label{int}
\end{equation}
where
\begin{equation}
p_{min}(x)=\left| \frac{M^2(1-x)^2-M_n^2}{2M(1-x)} \right|.
\label{pmin}
\end{equation}
We observe now that
$p_{min}$ remains unchanged
if we replace $x$ by $2 -x$
\begin{equation}
p_{min}(x) =p_{min}(2-x)\,.
\end{equation}
But $p_{min}(x)$ is not the only
source of $x$-dependence in  $q(x)$. After
exploiting the $delta$-function as in (\ref{int}),
the matrix elements appearing in (\ref{qxp})
also depend on $x$. Hence Eq.~(\ref{qxp}) is in general non invariant
under the substitution $x \to 2-x$. However,
  in the large-$x$
limit (which, according to (\ref{pmin}), is equivalent to the
large-$\vert { \Vec p}_n\vert $ limit) the matrix elements in (\ref{qxp})
tend to become
 $x$-independent. The reason is simple. If we describe
the quarks inside the hadron by Dirac spinors with
an upper component $u(\vert {\Vec p}_n \vert)$
and a lower component $v(\vert {\Vec p}_n \vert)$, the
$x$-dependence of the matrix
elements is contained in  interference terms
of the type $u(\vert {\Vec p}_n \vert) v(\vert {\Vec p}_n \vert)$
-- see \cite{noi}. Now
for large momenta $u$ and $v$ must behave as plane waves
and their product vanishes when integrated over
$\vert {\Vec p}_n \vert$.
Therefore as $x$ gets large the quark distribution
tends to become invariant with respect to the substitution
$x \to 2 -x$, namely $q(x) \simeq q(2-x)$, or
equivalently
\begin{equation}
q (1/z) \simeq q(2 - 1/z),
\label{qonez}
\end{equation}
in the large-$z$ limit.
Incidentally, we notice that the same happens in the limit
where relativistic effects can be neglected.
By combining Eq.~(\ref{ndly}) with
Eq.~(\ref{qonez}) we finally get the relation we have
anticipated above
\begin{equation}
\frac{1}{z} \, D(z) \simeq q(2-1/z).
\label{NR}
\end{equation}
Notice that for $z \ge 0.5$ this relation connects
(approximately) the {\em physical region} of DIS to
the {\em physical region} of IA.
Eq.~(\ref{NR}) is intended to be valid at a
fixed and small scale $\mu^2 < 1$ GeV$^2$.

\begin{figure}[htb]
\begin{center}
\leavevmode {\epsfysize=4.5cm \epsffile{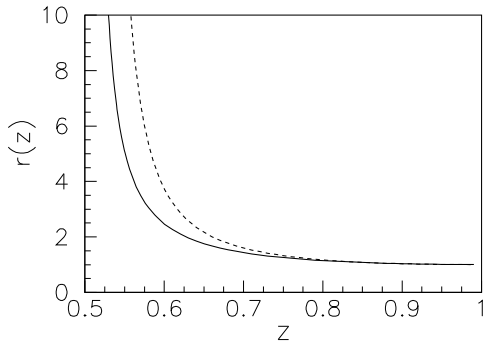}}
\end{center}
\caption[*]{\small \baselineskip=10pt
The ratio $r(z)=q(1/z)/q(2-1/z)$
in the light-cone quark model.
The solid and dashed curves are the results
in the light-cone quark model for the Gaussian type
wavefunction
(\ref{BHL}) and the power-law type wavefunction (\ref{PL}),
with $m_q=220$ MeV, $\beta=450$ MeV, $m_D=800$ MeV, and
$a=3.5$.
 }
\label{dm1f1}
\end{figure}

We can check  the validity of (\ref{qonez}) by an explicit model
calculation.
We use  a quark-diquark model
\cite{Ma96} in the framework of the
light-cone approach to quark distribution functions
\cite{Bro82,Ma86}. In this model the probability
to hit a quark
of mass $m_q$ and transverse momentum $\Vec{k}_{\perp}$
inside the nucleon, leaving a
 spectator-diquark
with mass $m_D$ in the state $D$, is
$q_D(x) \sim \int {\rm d}^2 {\Vec k}_{\perp}
\, |\varphi_D(x,{\Vec k}_{\perp})|^2$,
where $\varphi_D(x,{\Vec k}_{\perp})$ is the momentum space
wave function of the quark-diquark system with invariant
mass ${\cal M}^2=\frac{m_q^2+ {\Vec k}_{\perp}^2}{x}
+\frac{m_D^2+{\Vec k}_{\perp}^2}{1-x}$.
For the light-cone wave function
$\varphi_D(x,{ \Vec k}_{\perp})$
we use two different
forms:
 the gaussian type wave function of the Brodsky-Huang-Lepage
model \cite{Bro82} and a power-law type wave function
\begin{equation}
\varphi_D(x,{\Vec k}_{\perp})=A_{\mathrm{BHL}}
\exp(-{\cal M}^2/8\beta^2),
\label{BHL}
\end{equation}
\begin{equation}
\varphi_D(x,{\Vec k}_{\perp})=A_{\mathrm{PL}}(1+{\cal M}^2/\beta^2)^{-a}.
\label{PL}
\end{equation}
In Fig.~\ref{dm1f1} we plot the
ratio
\begin{equation}
r(z)=\frac{q(1/z)}{q(2-1/z)}\;.
\end{equation}
One can notice that
$r(z)$ approaches 1 at large $z$, as we expected,  and that
the two model wavefunctions lead to very similar results.
The sharp increase
of $r(z)$ as $z \to 0.5$ is due to the vanishing
of the denominator $q(2 -1/z)$ when its argument tends to zero.
This is an artifact of the quark-diquark
light-cone model which
is purely valence-like. In more sophisticated models
containing a  sea of quarks and antiquarks $q(x)$
does not vanish as $x \to 0$ and the increase of
$r(z)$ is tamed so that no spurious singularity exists $z =0.5$.

Let us come now to the phenomenology of the
new relation (\ref{NR}).
If we stick to leading order QCD and use
Eq.~(\ref{GL}), we can translate Eq.~(\ref{NR}) in terms
of structure functions as (remember that at large $x$
or $z$ the evolution is dominated by the quark
splitting functions)
\begin{eqnarray}
\bar{F}_1(z) &=& 3 F_1(2-1/z);
\label{NRF1} \\
z \bar{F}_2(z)&=&- \frac{3}{2 -1/z} F_2(2-1/z).
\label{NRF2}
\end{eqnarray}
Using standard parametrizations for the DIS structure
functions we can predict the IA structure functions
at large $z$ by means of Eqs.~(\ref{NRF1},\ref{NRF2}).
A caveat is in order. Since there are only few DIS experimental
data for $x > 0.7$, the quark distributions in this region
are not very well known. This introduces some
uncertainty in our predictions.

\begin{figure}[htb]
\begin{center}
\leavevmode {\epsfysize=4.5cm \epsffile{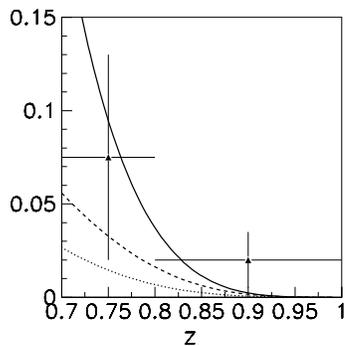}}
\end{center}
\caption[*]{\small \baselineskip=10pt
The structure function $z^3\bar{F}_2(z)$ in $e^+e^-$ annihilation.
The data are the experimental results from DASP
at $Q^2=13$ GeV$^2$ \cite{DASP,Kon93}.
The solid curve is the prediction based on (\ref{NR}) and
(\ref{NRF2}). The dotted curve is the prediction
of the traditional Gribov-Lipatov relation (\ref{GLRF}). The dashed
curve is the prediction based on
(\ref{GLRR}).
For the quark distribution functions
we used the CTEQ parametrization \cite{CTEQ5}.
 }
\label{dm1f2}
\end{figure}

 In Fig.~\ref{dm1f2} we compare the DASP data on $z^3\bar{F}_2(z)$
with the predictions based on the new relation
(\ref{NRF2}) and on the traditional Gribov-Lipatov
relation (\ref{GLRF}). For comparison,
we also show the results for $z^3\bar{F}_2(z)$ based
on the approximation (\ref{GLRR}).
For the DIS structure functions we used the CTEQ5L
parametrization \cite{CTEQ5}.
We find that the result of the new relation (\ref{NRF2})
is in better agreement with the data at $z \to 1$.
Clearly, precision measurements of
both $\bar{F}_2(z)$ and $F_2(x)$ at large $z$ and $x$
would allow a more conclusive check of (\ref{NRF2}).

We already pointed out that Eqs.~(\ref{NRF1},\ref{NRF2})
are valid
at leading order only. At next-to-leading order the evolution
of nonsinglet distribution and fragmentation functions is
different \cite{curci,bluemlein}.
Since the fragmentation functions evolve more  rapidly,
NLO effects lead to
 a suppression
at large $z$ of the LO results for $\bar{ F}_2$
shown in Fig.~\ref{dm1f2}.
Due to the large uncertainty
of the present-day data, in this paper we chose for simplicity to
stick to a leading-order phenomenological treatment. Using NLO
splitting functions for the fragmentation functions \cite{Flo98b}
and the new relation (\ref{NR}) as the initial
condition for the evolution one can
calculate the NLO corrections to Eqs.~(\ref{NRF1},\ref{NRF2}).

In conclusion, we presented a new relation
between distribution and fragmentation functions in their
physical regions, which leads
to simple testable relations between DIS and IA
structure functions. A revised form of Gribov-Lipatov
relation with a color factor and an additional
factor of $z$ is also proved to be an approximate
relation at large $z$.
An immediate application of the new relation
connecting $q(x)$ to $D(z)$ is in the study of the
$\Lambda$ polarization near the $Z$ resonance
in $e^+e^-$ annihilation and in polarized lepton DIS
scattering.
Using the
Gribov-Lipatov relation and the QCD
counting rules for the quark helicity distributions
\cite{Bro95}, it was found that the data
are not satisfactorily reproduced
\cite{Ma00}.
We have checked that the situation improves
 if the
new relation (\ref{NR}) is used.

\vspace{0.5cm}

\noindent{\bf Acknowledgments: } We would like to thank
S.J.~Brodsky, L.~Caneschi,  R.L.~Jaffe, L.N.~Lipatov, J.P.~Ma, and
I. Schmidt for helpful discussions. One of the authors (B.-Q. M.)
is grateful to INFN, Sezione di Ferrara, for hospitality and
partial support. This work is partially supported by National
Natural Science Foundation of China under Grants No.~19605006 and
No.~19975052.

\end{document}